\documentclass[prb,twocolumn,amsmath,amssymb,amsfonts]{revtex4}
\usepackage[pdftex]{graphicx}
\pdfoutput=1
\usepackage{color}

\newcommand{\ud}{\mathrm{d}}
\newcommand{\uprime}{^{\prime}}

\newcommand{\subhigh}{\mathcal{A}}
\newcommand{\sublow}{\mathcal{B}}

\begin{document}

\title{Carrier multiplication yields in CdSe and CdTe nanocrystals
by transient photoluminescence}

\author{Gautham Nair}
\author{Moungi G Bawendi}
\email{mgb@mit.edu}
\affiliation{Department of Chemistry, Massachusetts Institute of Technology, 77 Massachusetts Avenue, Cambridge, Massachusetts 02139}

\begin{abstract}
	Engineering semiconductors to enhance carrier multiplication (CM) could lead to increased photovoltaic cell performance and a significant widening of the materials range suitable for future solar technologies. Semiconductor nanocrystals (NCs) have been proposed as a favourable structure for CM enhancement, and recent measurements by transient absorption have shown evidence for highly efficient CM in lead chalcogenide and CdSe NCs. We report here an assessment of CM yields in CdSe and CdTe NCs by a quantitative analysis of biexciton and exciton signatures in transient photoluminescence decays. Although the technique is particularly sensitive due to enhanced biexciton radiative rates relative to the exciton, $k^{rad}_{BX} > 2 k^{rad}_X$, we find no evidence for CM in CdSe and CdTe NCs up to photon energies $\hbar \omega > 3 E_g$, well above previously reported relative energy thresholds.
\end{abstract}


\maketitle
	Carrier multiplication (CM) in the form of impact ionization is a well-understood phenomenon in bulk semiconductors \cite{WolfJAP98,HarrisonJAP99}. The process, consisting of inelastic scattering of energetic charge carriers and valence electrons to create additional $e$-$h$ pairs, normally has high energy thresholds and low efficiency due to momentum conservation requirements and competition from ultrafast intraband relaxation. While this conventional bulk CM has found a particular application in avalanche photodiodes for single photon detection, efficient CM following optical excitation could have a significantly wider impact in the area of solar energy conversion. 
	
	In a typical photovoltaic cell with a single active layer, photon energy in excess of the bandgap is lost by rapid thermalization. The CM process, if efficient, could harvest this excess energy into additional $e$-$h$ pairs, boosting the maximum theoretical power conversion efficiency from 32\% to $>$40\% \cite{ShockleyJAP61,KlimovAPL06}, and, more importantly, widening the range of candidate materials for new solar technologies to include previously ignored narrow-gap semiconductors. Strongly confined semiconductor nanocrystals (NCs) have been proposed as candidate structures for efficient CM \cite{NozikARPC01} because of an anticipated relaxation of momentum conservation constraints \cite{ChepicJL90,WangPRL03} and potential slowing of competing phonon-mediated intraband cooling due to the discrete electronic structure (``phonon bottleneck'') \cite{NozikARPC01}. 

	Recently, transient absorption (TA) measurements have shown evidence of efficient ultrafast CM in IR-emitting PbSe, PbS, and PbTe NCs above a photon energy threshold $\hbar \omega = 3E_g$ \cite{SchallerPRL04,EllingsonNL05,MurphyJACS06}. In extensions of the work, subsequent TA measurements have indicated that a single-photon could generate up to 7 $e$-$h$ pairs in PbSe \cite{SchallerNATPHYS05} and that CM is similarly efficient in visible-emitting CdSe NCs above a $\hbar \omega = 2.5E_g$ threshold \cite{SchallerAPL05}.

	The conclusions of the TA measurements suggest new and unique underlying physics as well as some interesting questions. First-principles theories explaining the balance of Coulomb coupling and phonon relaxation rates implied by the experiments have yet to emerge. At the same time, studies on intraband relaxation in CdSe and PbSe NCs at room temperature have found fast cooling dynamics that do not appear consistent with a phonon bottleneck \cite{Guyot-SionnestJCP05,WehrenbergJPCB02,HarboldPRB05}. In addition, some aspects of the experimental data are intriguing, such as similar CM effects seen in PbSe and CdSe despite the very different state structures at threshold, and the observed linear dependence of CM yields on excess energy \cite{SchallerAPL05}. 
	
	
	We present here an assessment of CM yields in CdSe and CdTe nanocrystals by transient photoluminescence (tPL). While complementary to TA in some ways, tPL is a background-free measurement better suited to the low excitation fluences necessary in CM studies. It is also more selective since it relates to the number of \emph{e}-\emph{h} pairs instead of single-particle state filling. Both TA and tPL capitalize on the unique, fast dynamics of multiexciton (MX) states \cite{KlimovSCIENCE00auger,CarugePRB04,BonatiPRB05} to isolate and quantify MX populations, and though more complicated to interpret, tPL becomes a useful and particularly sensitive technique when carefully calibrated. A recent study has analysed CM yields in CdSe NCs by tPL and concludes that there is agreement with previous TA determinations \cite{SchallerJPCB06}. Their analysis and experiment differs from ours in several significant ways. We find instead that CM efficiency in CdSe and CdTe NCs is close to zero even for photon energies up to $3.1 E_g$. Implications are discussed.

	Tunable UV excitation pulses were generated by nonlinear mixing of the visible output and 3.1 eV remnant of an optical parametric amplifier (Coherent OPA 9400) pumped by a 250 kHz amplified Ti:sapph laser (Coherent RegA 9000). In a seperate experiment, the Ti:sapph was tuned to the red and 5.9 eV pulses were obtained by doubling its second harmonic. Room temperature hexane dispersions of NCs in 1mm-path length cuvettes were excited at $45^{\circ}$ incidence. Emission was collected front-face and spectrally dispersed onto a streak camera (Hamamatsu C5680). 
We studied both organic ligand-capped CdSe or CdTe (core) and CdSe/ZnS or CdSe/ZnCdS (core/shell) overcoated particles \cite{EPAPS}. 

\begin{figure}
\includegraphics*[scale=1.0,viewport=6 0 252 112]{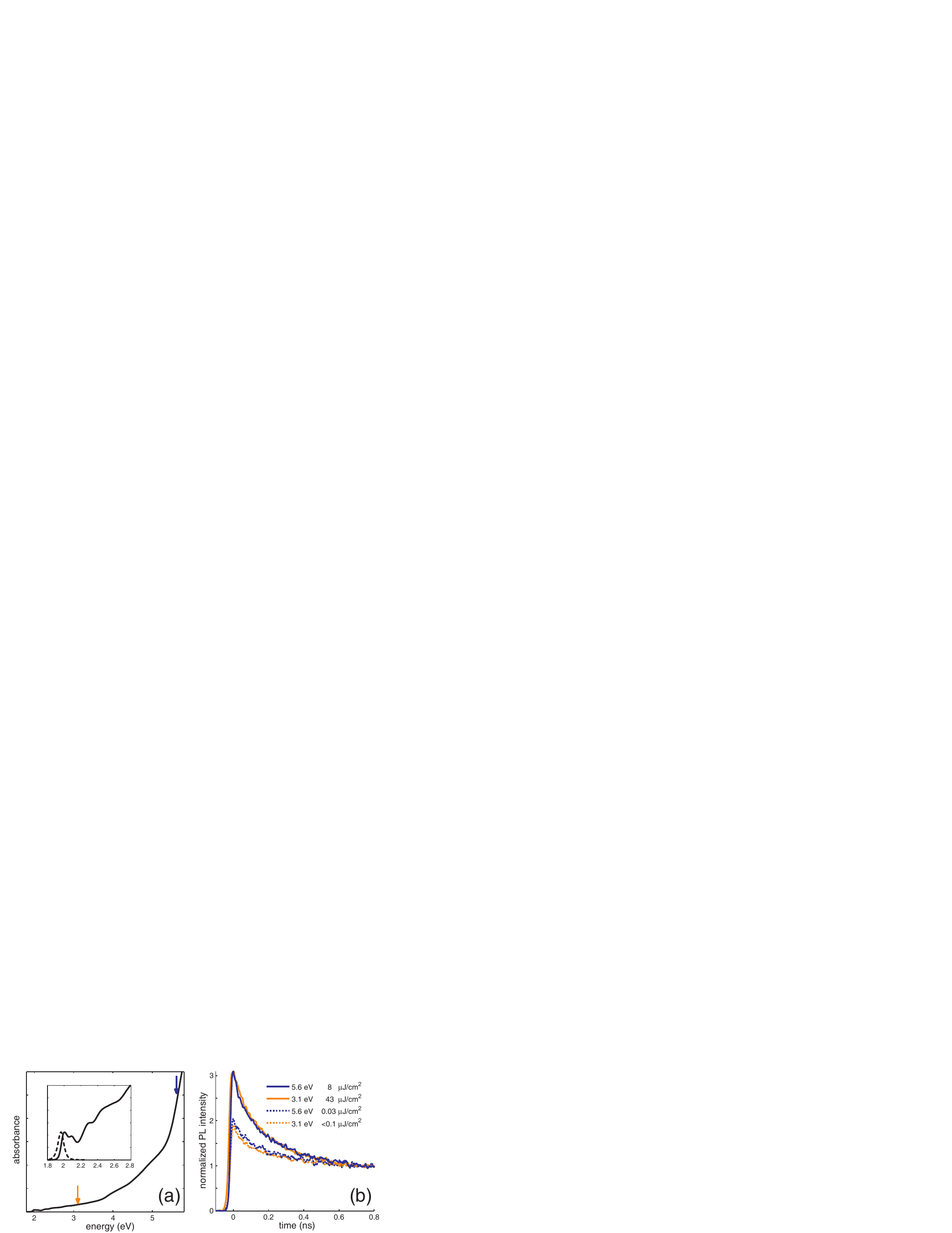}
\caption{\label{fig_basics} (color online). (a) Absorption spectrum of a typical CdSe NC core sample used. Excitation energies employed in tPL are indicated by the arrows (inset) Detail of band edge absorption structure and emission spectrum. (b) Band edge luminescence dynamics of the sample at the two indicated excitation energies for peak pulse fluence as noted.} 
\end{figure}

	Fig.\ref{fig_basics} shows PL decays of a representative sample of $E_g$=2.0 eV CdSe NCs under weak and strong excitation at 3.1 eV and 5.6 eV ($E_g$ determined from the lowest absorption feature). At high fluence both decays show an additional fast component consistent with biexciton (BX) emission. 
	Remarkably, and unlike data from Ref.\onlinecite{SchallerJPCB06} which shows a fast component under UV excitation, we find that the two low-fluence decays follow each other closely, suggesting that CM is less efficient than previously reported.
	 A quantitative determination of the CM yield requires first a careful characterization of the BX tPL signature, which we describe in detail.  

\begin{figure}
\includegraphics*[scale=1.0,viewport=0 0 240 210]{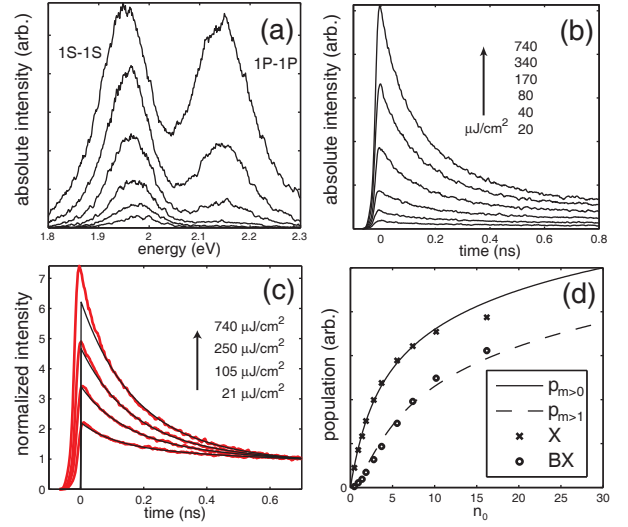}
\caption{\label{fig_400summary} (color online). (a) Transient PL spectra of the NCs in Fig.1 collected from $t=-10 \text{ ps}$ to $t=10\text{ ps}$ and (b) decays integrated from $1.89$ to $2.02$ eV under 3.1eV excitation with a peak fluences as noted. (c) A separate subset of our data, showing PL decays normalized at long times. The black lines are fits to the form $a_{BX}e^{-t/\tau_{BX}}+a_{X}f(t)$ where $\tau_{BX}=185\text{ ps}$ and $f(t)$, the single X dynamics, are kept constant \cite{FOOTNOTE_NOFASTBX_FITDETAILS, EPAPS}. (d) (lines) Predicted initial NC populations in an X or higher state ($p_{m>0}$) or in a BX or higher state ($p_{m>1}$) plotted against the peak average number of $e$-$h$ pairs created, $n_0=\textrm{max}\{n(\vec{r})\}$. ($\times$,$\circ$) Scaled and fit $a_X$ and $a_{BX}$ PL components extracted from fits to measured decays.}
\end{figure}

	Results of an excitation-fluence series at 3.1 eV are shown in Fig \ref{fig_400summary}. Decays at low fluence are dominated by exciton (X) emission. In most cases, X dynamics are multiexponential with contributions from NCs with different non-radiative relaxation and trapping rates \cite{SchlegelPRL02,FisherJPCB04}. The decays are adequately described by a biexponential, $f(t)=\exp{\left(-t/\tau_{X slow}\right)}+c_{fast} \exp{\left(-t/\tau_{X fast}\right)}$, chosen for simplicity, where
 $\tau_{X fast}\approx100\text{-}300\text{ ps}$ and $\tau_{X slow}\approx 1\text{-}10\text{ ns}$. With increasing fluence, early-time spectra show biexciton (BX) emission at the band edge and a further, blue-shifted feature corresponding to 1P-1P emission from higher multiexcitons (Fig.\ref{fig_400summary}a \& \ref{fig_400summary}b) \cite{CarugePRB04,BonatiPRB05}. The BX state then decays quickly with a size-dependant lifetime $\tau_{BX}\approx 0.1\text{-}1\text{ ns}$ due to a fast non-radiative ``Auger''-like coulomb process  \cite{KlimovSCIENCE00auger}. As shown in Fig.\ref{fig_400summary}c, the measured tPL decays are well described by a superposition of X dynamics and an additional single exponential BX component, $ a_{BX}\exp{\left(-t/\tau_{BX}\right)}+a_{X} f(t)$ 
 %
 \cite{FOOTNOTE_NOFASTBX_FITDETAILS}.

	The relationship between the observed tPL decay amplitudes, $a_X$ and $a_{BX}$, and the underlying MX populations was studied using a first-order kinetic model of MX relaxation, $p_m\uprime(t)=-k_m p_m + k_{m+1} p_{m+1}$, giving $a_{X} \propto p_{m>0}^o k^{rad}_{X}$ and
\begin{equation}
a_{BX} \propto p_{m>1}^o \left(k^{rad}_{BX}-k^{rad}_X\right)  \left(1+\frac{k_2}{k_3-k_2}\frac{p_{m>2}^o}{p_{m>1}^o}+\ldots\right) 
\nonumber
\end{equation}
where $p_m$ is the relative population of NCs with $m$ electrons and holes, $p_m^o$ are initial values at t=0, $k_m$ are the MX decay rates, and $k^{rad}_X$ and $k^{rad}_{BX}$ are the X and BX radiative rates. The $a_{BX}$ term includes BX populations formed by cascaded decay of higher MX states and is proportional to $k_{BX}^{rad}-k_X^{rad}$ since BX emission is partially offset by correspondingly reduced X emission at early times \cite{EPAPS}.
The $p_m^o$ are related to the incident laser power assuming poissonian photon absorption statistics and explicitly accounting for the inhomogeneous excitation profile of the beam:
	\begin{equation}
	 p_m^o=\int{ \frac{n(\vec{r})^m}{m!} e^{-n(\vec{r})} \ud^3 \vec{r}} \qquad n(\vec{r})=j_p(\vec{r})\sigma \nonumber
	\end{equation} 
where $j_p(\vec{r})$ is the measured photon flux at $\vec{r}$, $n(\vec{r})$ is the average number of absorbed photons per NC, and $\sigma$ is the absorption cross-section, treated here as an adjustable parameter. 

Band edge PL decays show growth and slow saturation of $a_X$ and $a_{BX}$ that fit reasonably well to the expected curves (see Fig.\ref{fig_400summary}d)    
%
\footnote{For our approximately gaussian beam profiles, $p^o_{m>0}$ transitions from $\sim{n_0}$ to $\sim\ln n_0$ instead of fully saturating, and the ratio $p^o_{m>1}/p^o_{m>0}$ is much smaller and more slowly approaches unity than for uniform illumination. Our fits neglect the last term in the expression for $a_{BX}$ which accounts for the small delaying effect of cascade from higher multiexcitons ($k>2$). This approximation leads to at most a $\approx 25\%$ overestimate of $(a_{BX}/a_X)_{sat}$ 
 and has been taken into consideration in our final CM analysis \cite{EPAPS}.}.
%
%
From the fits we extract the $a_{BX}/a_X$ ratio expected at BX saturation ($p^o_{m=2}=p^o_{m>0}$) and find a sample-dependant $(a_{BX}/a_X)_{sat}$ value in the range $3$-$6$. This implies a substantially faster radiative rate of the BX relative to the X and leads to enhanced sensitivity of tPL for detection of small multiexciton populations, as is illustrated by the prominence of BX features in Fig.\ref{fig_400summary}c. We estimate $k^{rad}_{BX}/k^{rad}_{X}>3\text{-}5$, consistent with previous measurements on high quality NCs showing $k^{rad}_{BX}/k^{rad}_{X} \approx 3$ \cite{FisherPRL05}.


	The observed faster radiative lifetime of the BX relative to the X is explained by the electronic fine structure of these states. Because emission from the X ground state is spin-forbidden, X luminescence is relatively slow ($k^{rad}_X \approx 0.05$ $\textrm{ns}^{-1}$) and consists mostly of emission from thermally populated spin-allowed bright states \cite{NirmalPRL95,EfrosPRB96,ShumwayPRB01}. However, transitions from the BX ground state to some states in the X fine structure are predicted to be optically allowed \cite{ShumwayPRB01}, indicating that $k^{rad}_{BX}$ can be significantly larger than the $2k^{rad}_X$ value one would predict from a simple carrier counting argument. 
	
	
	
\begin{figure}
\includegraphics*[scale=1.0, viewport=0 0 204 204]{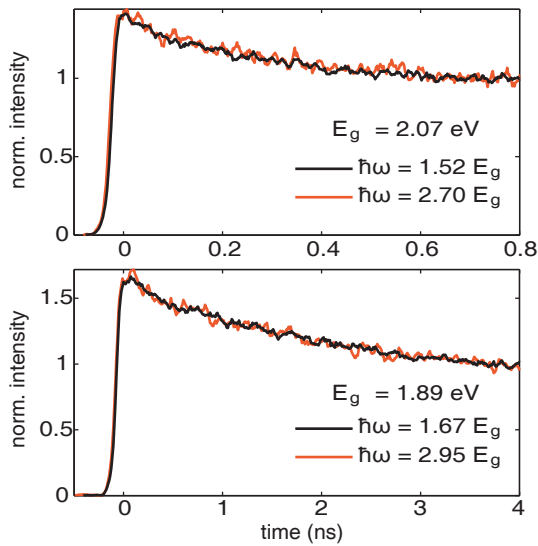}
\caption{\label{fig_nocm} (color online). Band edge PL decays of $E_g$=$2.07$ eV (top) and $E_g$=$1.89$ eV (bottom) core/shell CdSe NCs under weak ($n_0 < 0.01$) excitation at 3.1eV (black) and 5.6 eV (red). The 5.6 eV excitation corresponds to $\hbar \omega$=$2.70 E_g$ and $2.95E_g$ respectively, for which TA measurements at 6.2 eV predict CM yields of 22\% and 50\% \cite{SchallerAPL05}}
\end{figure}
	
\begin{figure}
\includegraphics*[scale=1.0,viewport=0 0 228 204]{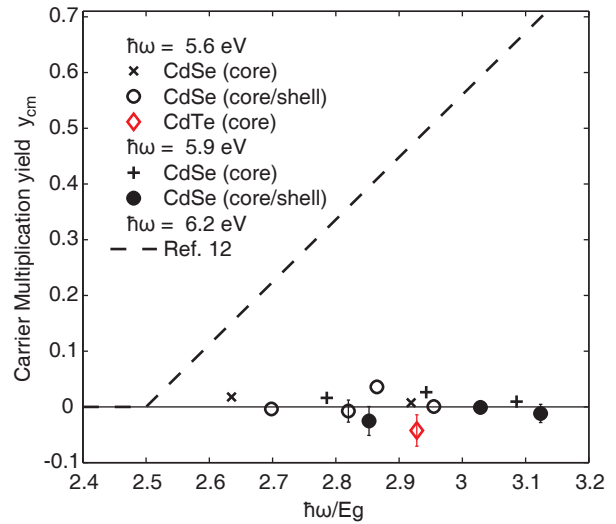}
\caption{\label{fig_cmvsE} (color online). Carrier multiplication yields under 5.6 and 5.9 eV excitation extracted from tPL decays, $y_{cm}=\frac{a_{BX}/a_X}{\left(a_{BX}/a_X\right)_{sat}}$. Conservative (small) values $\left(a_{BX}/a_X\right)_{sat}$ were used for each sample. $2\sigma$-wide error bars from repeated measurements are shown if larger than the symbols. The dashed line is the $y_{cm}$ vs. $\hbar\omega/E_g$ dependence found in the TA studies of Ref.\onlinecite{SchallerAPL05} using $\hbar\omega=6.2$eV.}
\end{figure}

	The sensitivity of the tPL experiment was thus exploited to estimate carrier multiplication yields in NC samples under UV excitation 
	\footnote{Concentrated (OD $\sim$1 at 3.1eV), stirred samples and short integration times were employed to minimize exposure at 5.6eV. Prolonged exposure leads to irreversible degradation reflected in emission quenching and faster, more strongly multiexponential X dynamics, which were monitored throughout the experiment to verify sample integrity.}.
	We find that, while signatures of multiexciton emission appear at high fluence, decays under weak 5.6 or 5.9eV excitation are close to indistinguishable from decays under weak 3.1eV (Fig. \ref{fig_nocm}) even for large NCs where the excitation energy $\hbar \omega$ exceeds $3E_g$. This contrasts with the measurements of Ref. \onlinecite{SchallerJPCB06} which show an additional fast component under UV excitation. Quantitative values shown in Fig.\ref{fig_cmvsE} of the CM yield, the fraction of photo-excited NCs initially found in the BX state, $y_{cm}=\frac{p^o_2}{p^o_1+p^o_2}$ 
	\footnote{It has been common in the literature to report CM yields in terms of an exciton internal quantum efficiency (IQE), which is related to our $y_{cm}$ by $\textrm{IQE}=100\% (y_{cm}+1)$.}
, were obtained as the ratio $y_{cm}=\frac{a_{BX}/a_X}{\left(a_{BX}/a_X\right)_{sat}}$ from fitting the UV-excited decays to the form $ a_{BX}\exp{\left(-t/\tau_{BX}\right)}+a_{X} f(t)$ \cite{EPAPS}.
	
		As seen in Fig.\ref{fig_cmvsE}, our results do not match the $\hbar\omega/E_g$ dependence of $y_{cm}$ found in TA measurements on CdSe NCs using 6.2 eV  excitation \cite{SchallerAPL05} (Fig.\ref{fig_cmvsE}). This could be because $y_{cm}$ is not only a function of the ratio $\hbar \omega/E_g$, as suggested by TA on PbSe NCs \cite{SchallerPRL04}, but also depends explicitly on $\hbar \omega$. However, our interpretation of recently published tPL data at 6.2 eV \cite{SchallerJPCB06} using our estimated $k^{rad}_{BX} > 3k^{rad}_X$ suggests a $y_{cm}$ of at most $\approx35\%$ even at $\hbar\omega/E_g\approx 3.2$, instead of the $70\%$ CM yield reported, which assumed $k^{rad}_{BX} = 2k^{rad}_X$ \cite{SchallerJPCB06}. tPL and TA assessments of CM yield in CdSe NCs therefore appear to disagree, and it is possible that CM in semiconductor NCs is generally less efficient and not as universal as has been thought.
	
	
	There are several reasons to expect a different CM enhancement in II-VI semiconductors compared to the lead chalcogenides. The electronic state structure in wide gap NCs at energies $\geq 1E_g$ in excess of the band edge is likely bulk-like with level-spacings that are small. On the other hand, the analogous electronic states in the lead chalcogenides might be more discrete in character because of the smaller $E_g$ and lighter effective masses. For these reasons, our experimental conclusions cannot be extended to the narrow band NCs, but the findings suggest the need to verify TA assessments of $y_{cm}$ in the lead chalcogenides with other techniques such as tPL. 

	We note that the nature of a potential CM enhancement mechanism in NCs is itself not well understood. Although theoretical schemes have made progress towards explaining some of the reported phenomenology, these calculations have yet to quantitatively account for the balance of Coulomb and intraband relaxation processes leading to CM enhancement in a way that is clearly consistent with what is already known about carrier cooling and Auger rates in NCs.
		Calculations based on a traditional impact ionization model \cite{FranceschettiNL06} reconcile fast CM ($< 1$ ps) with the slower band-edge Auger multiexciton relaxation ($\sim 100$ ps) but make the assumption of constant Coulomb matrix elements between X and BX subspaces. While this ``random-$k$''-like approximation appears justified for the specific case of bulk Si at high energies \cite{KanePR67,WolfJAP98}, it might not apply to direct-gap semiconductor NCs, especially since the NC surface is thought to play a large role in Auger-like and other Coulomb processes \cite{WangPRL03}. Other researchers have pursued a generalized treatment, going beyond the perturbative approach of typical impact ionization calculations to examine the effect of phase and population relaxation rates on the CM process \cite{EllingsonNL05,ShabaevNL06}. Calculations on PbSe NCs using a simple model of the electronic structure have shown efficient CM \cite{ShabaevNL06} but, in doing so, obtain large values of Coulomb matrix elements that appear to conflict with the much slower observed Auger rates. The third approach, virtual-state-mediated CM  \cite{SchallerNATPHYS05}, has so far not taken dephasing or population decay into account but could be generalized with the introduction of decay rates in the energy denominators of their second order perturbation expression ($E \rightarrow E-i\Gamma$). All theoretical approaches so far thus hinge on the relative rates of coulomb interaction and intraband relaxation at threshold, but neither has been measured or accurately calculated for NCs. 
		
	In summary, we have determined CM efficiencies in CdSe and CdTe NCs by transient photoluminescence. Exciton and biexciton features were first characterized under 3.1 eV excitation from which we find a relatively fast BX radiative rate $k^{rad}_{BX} > 3k^{rad}_{X}$. Measurements under weak 5.6 and 5.9 eV excitation show no evidence of biexciton generation, and thus no CM, up to photon energies as high as $3.1E_g$.


\begin{appendix}


\section{Experimental methods}

\subsection{Sample preparation}
	Semiconductor nanocrystals (NCs) were synthesized by high temperature pyrolysis of precursors \cite{FisherJPCB04}. The product was purified by precipitation once with butanol and methanol and redispersed in hexane. Samples were diluted to an appropriate concentration and introduced in 1mm path length quartz cuvettes. For excitation-fluence studies, dilute solutions with an optical density (OD) of $\approx 0.1$ at 3.1eV were used, but for measurements under UV we employed very concentrated samples (OD $\approx 1$ at 3.1eV) to minimize the effects of degradation. A small magnetic bar was added to the solution. It was positioned near the excitation point and rotated vigorously in a plane parallel to the cuvette surface using an external magnetic stir plate. All measurements and manipulations were carried out at room temperature.

\begin{figure}
\centering
\includegraphics*[scale=1.0,viewport=0 0 240 264]{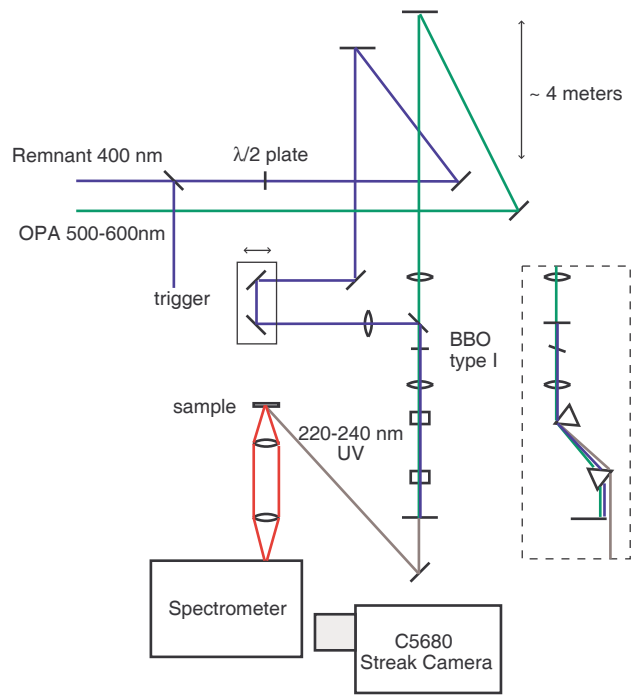}
\caption{\label{FIG_setup} Apparatus used for studying tPL of NCs with excitation energies up to 5.6eV}
\end{figure}

\subsection{tPL apparatus and measurement methodology}
	The experimental setup used for studying CM by tPL with 5.6eV excitation is shown in Fig.\ref{FIG_setup}. We isolated the 5.6 eV sum frequency generated beam from the 2.5 eV and 3.1 eV fundamental beams using a prism pair. This was confirmed by checking that UV-excited tPL signals vanish if the relative delay of the fundamental beams is adjusted or if a UV absorber (such as a thin glass coverslip) is placed in the beam path. 
	For excitation at 3.1 eV, the OPA was blocked and the 3.1 eV beam was directed onto the same excitation spot on the sample. For excitation at 5.9 eV, the Ti:sapph was tuned to the red and its second harmonic was directly doubled in the same nonlinear crystal. The resulting SHG was weak but sufficiently strong for tPL collection.
	
		The excitation beams were characterized spectrally with a fiber spectrometer (Ocean Optics QE65000) and their spatial profiles were obtained by directly imaging onto a CCD camera (Roper Scientific MicroMax) placed at the sample position.
	
	CM determination measurements consisted of alternating acquisitions under weak 3.1 eV and UV excitation. Exposure to UV was minimized by using low excitation power ($\leq 50 \mu$W) and short integration times ($\sim$ 5 min). Under these conditions, tPL decays excited with weak 3.1 eV remained unchanged throughout the length of the experiment, confirming that sample integrity was not compromised.

\section{Dynamics of multiexciton PL}

\subsection{Kinetic model}

We analyzed the experimentally observed PL dynamics with a first order kinetic model of multiexciton relaxation:
\begin{eqnarray}
\frac{\ud p_1(t)}{\ud t}&=&-k_1 p_1(t)+k_2 p_2(t) \nonumber \\
\frac{\ud p_2(t)}{\ud t}&=&-k_2 p_2(t)+k_3 p_3(t) \nonumber \\
&\vdots& \nonumber  \\
\frac{\ud p_j(t)}{\ud t}&=&-k_j p_j(t)+k_{j+1} p_{j+1}(t) \label{EQNfirstorder}
\end{eqnarray}
where $p_j(t)$ are the relative populations of NCs in the $j$-th multiexcitonic state (i.e. $j$ electrons and holes), and $k_j$ are the state decay rates.  The transient photoluminescence (tPL) signal seen on the detector will have contributions from all populated states. We are particularly interested in the dynamics of emission from the $1S_e-1S_h$ transition (i.e. band-edge luminescence), which is denoted by $s(t)$ and given by:

\begin{equation}
\label{EQN_soft}
s(t)\propto k_1^{rad} p_1(t)+k_2^{rad} p_2(t)+\ldots=\sum_{j} k_j^{rad} p_j(t)
\end{equation}  
where $k_j^{rad}$ is the rate of radiation of the $j$-th multiexciton state from the $1S$-$1S$ level, which in the case of $j=1$ and $j=2$ corresponds to the total radiative decay rates of the X and BX, $k_X^{rad}$ and $k_{BX}^{rad}$, respectively.

\subsection{Solution of the rate equations}

For clarity, first we solve the model assuming an initial population of only X, BX and TX states (i.e. $p_j(t=0)=0$ for $j>3$). We find:
\begin{eqnarray}
p_3(t)&=&p_3^oe^{-k_3 t} \nonumber \\
p_2(t)&=&\left(p_2^o +\frac{k_3}{k_3-k_2}p_3^o\right)e^{-k_2 t}-\frac{k_3}{k_3-k_2}e^{-k_3 t} \nonumber \\
p_1(t)&=&\left(p_1^o+\frac{k_2}{k_2-k_1}p_2^o+\frac{k_2}{k_2-k_1}\frac{k_3}{k_3-k_1}p_3^o\right)e^{-k_1 t} \nonumber \\
& &-\frac{k_2}{k_2-k_1}\left(p_2^o +\frac{k_3}{k_3-k_2}p_3^o\right)e^{-k_2 t}\nonumber \\
& &+\frac{k_2k_3}{(k_3-k_2)(k_3-k_1)}p_3^o e^{-k_3 t} \nonumber
\end{eqnarray}
where the $\{p_j^o\}$ denote initial populations at $t=0$. Substituting into Eqn. \ref{EQN_soft} and grouping terms according to their time dependence, we find the luminescence signal is given by
\begin{eqnarray}
s(t)&\propto&k_1^{rad}\left(p_1^o+\frac{k_2}{k_2-k_1}p_2^o+\frac{k_2}{k_2-k_1}\frac{k_3}{k_3-k_1}p_3^o\right)e^{-k_1 t} \nonumber \\
& &+\left(k_2^{rad}-k_1^{rad}\frac{k_2}{k_2-k_1}\right)\left(p_2^o +\frac{k_3}{k_3-k_2}p_3^o\right)e^{-k_2 t} \nonumber \\
& &+\left(\ldots\right)e^{-k_3 t} \nonumber
\end{eqnarray}

We are interested here only in the components of the luminescence with dynamics corresponding to the exciton and biexciton decay rates. The above result is further simplified using $\frac{k_1}{k_j} \approx 0$ for $j>1$ since the multiexciton Auger decay rates are more than an order of magnitude larger than X decay rates. The decomposition of $s(t)$ as a sum of exponentials is thus given by:
\begin{eqnarray}
e^{-k_1t}& : & a_X=k_1^{rad}\left(p_1^o+p_2^o+p_3^o\right) \nonumber \\
e^{-k_2t}& : & a_{BX}=(k_2^{rad}-k_1^{rad})\left(p_2^o+p_3^o\right)\times\nonumber\\
& &\qquad\left(1+\frac{k_2}{k_3-k_2}\frac{p_3^o}{p_2^o+p_3^o}\right) \nonumber \\
s(t)&=&a_{X}e^{-k_1 t}+a_{BX}e^{-k_2 t}+\left(\ldots\right)e^{-k_3 t} \nonumber
\end{eqnarray}

The $e^{-k_1 t}$ component of $s(t)$ is simply proportional to the X radiative rate multiplied by the population of NCs that start in an X state or higher, since MX states eventually decay to the X state. The BX component has three factors. It is firstly proportional to the difference of X and BX radiative rates because BX luminescence is partially offset by a dip in X luminescence, as it is still in the process of being populated. The second factor in the expression for $a_{BX}$ is proportional again to the sum of the initial populations in BX or higher states. The third term captures the small delaying effect of cascaded decay from the TX. In NCs originally in the TX state, there is a short $t\sim k_3^{-1}$ delay before population of the BX, which then decays normally at a rate $k_2$. Thus, if the $e^{-k_2 t}$ dependence of the subsequent BX decay is extrapolated back to $t=0$ one finds a slightly larger value of $a_{BX}$ than would be expected in the absence of the delay.  

\begin{figure}
\centering
\includegraphics*[scale=0.8,viewport=0 0 264 204]{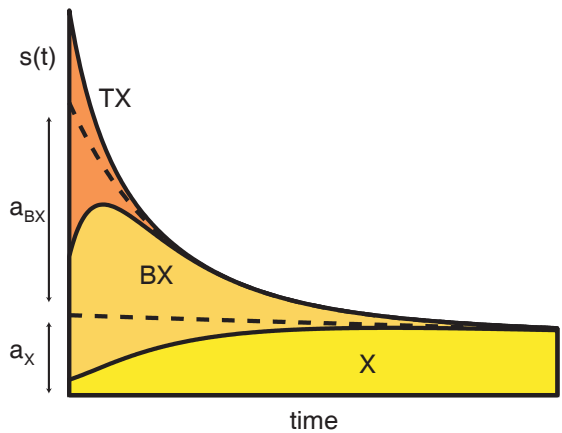}
\caption{\label{FIG_MXtPL} Decomposition of the tPL decay $s(t)$. Solid areas show emission by X,BX and TX states. In contrast, the dashed lines show the decomposition of $s(t)$ into a sum of exponentials. $s(t)=a_Xe^{-k_X t}+a_{BX}e^{-k_{BX} t}+(\ldots)e^{-k_{TX} t}$}
\end{figure}

Fig. \ref{FIG_MXtPL} serves to illustrate the contributions to $s(t)$ in a typical case where $p_1^o,\,p_2^o,\,p_3^o\neq 0$, as would be the case following pulsed excitation and assuming Poissonian photon absorption statistics. The solid areas show the decomposition of $s(t)$ according to the state involved in the emission  $s(t)=k_1^{rad}p_1(t)+k_2^{rad}p_2(t)+k_3^{rad}p_3(t)$. The dashed lines show the decomposition into a sum of exponentials, $s(t)=a_{X}e^{-k_1 t}+a_{BX}e^{-k_2 t}+\left(\ldots\right)e^{-k_3 t}$, as would be obtained from a multiexponential fit to a measured $s(t)$.

The expressions for $a_X$ and $a_{BX}$ derived above for the restricted case $p_j^o=0 \,\, \forall j>3$ suggest a generalization:
\begin{eqnarray}
a_X&=&k_1^{rad}\left(p_1^o+p_2^o+\ldots \right)=k_1^{rad}p_{j>0}^o \label{EQNax} \\
a_{BX}&=&(k_2^{rad}-k_1^{rad})p_{j>1}^o\times\nonumber\\
& &\quad\left(1+\frac{k_2}{k_3-k_2}\frac{p_{j>2}^o}{p_{j>1}^o}+\ldots\right) \label{EQNabx}
\end{eqnarray}
which is proved for the general case of arbitrary $\{p_j^o\}$ in the section \ref{SECTIONlaplace}.


\subsection{Effect of multiexponential X dynamics}

	The analysis so far has relied on the assumption of a homogeneous sample of NCs with identical decay rates. In fact, experimental tPL data often show a distribution of X decay rates. Considering that high quality core/shell samples do show very close to single exponential dynamics, we identify two NC subpopulations in each sample, $\subhigh$ and $\sublow$, corresponding to high and low quality NCs respectively. The distribution could be static or might involve switching between $\subhigh$ and $\sublow$ on slow timescales \cite{SchlegelPRL02, FisherJPCB04}. We denote by $p_j^\subhigh$ and $p_j^\sublow$ the relative number of NCs in each population in the $j$-th multiexcitonic state. 
		
	Our experimental data shows that, although the X decays can be significantly multiexponential, the additional contribution attributed to biexciton emission is well described by a \emph{single} exponential. This suggests that the low-quality subpopulation does not contribute to the BX signal within our experiment's time resolution. Rapid BX decay in those NCs could be explained by the same traps or nonradiative mechanisms involved in their fast X decay, but could also be due to Auger decay enhancement on rough or defective surfaces, as would be anticipated from theoretical treatments \cite{WangPRL03,EfrosNRL02}. For our purposes we can thus approximate the MX decay down to the X state in type-$\sublow$ NCs as instantaneous :
\begin{equation}
p_{j>1}^\sublow(t>0)\approx 0 \qquad \textrm{and} \qquad p_1^\sublow(t)=p_{j>0}^{o\,\sublow} \cdot h(t)
\end{equation}
where $h(t)$ is an exponential or multiexponential decay with $h(0)=1$. On the other hand, the $\subhigh$ population dynamics are single exponential and follow the treatment of the previous section. Assuming that the radiative rates of $\subhigh$ and $\sublow$ NCs are identical, we obtain an expression for the tPL signal as follows:
\begin{eqnarray}
s(t)&\propto&(k_2^{rad}-k_1^{rad})p_{j>1}^{o\,\subhigh}\left(1+\frac{k_2}{k_3-k_2}\frac{p_{j>2}^{o\,\subhigh}}{p_{j>1}^{o\,\subhigh}}+\ldots\right) e^{-k_2 t}\nonumber \\
& &+k_1^{rad}\left(p_{j>0}^{o\,\subhigh}\cdot e^{-k_1 t}+p_{j>0}^{o\,\sublow}\cdot h(t)\right)+\ldots \nonumber
\end{eqnarray}

To simplify, we introduce $b=p_{tot}^\sublow/p_{tot}^\subhigh$, the ratio of the total number of NCs of type $\sublow$ to type $\subhigh$. Since photon absorption is not expected to be affected by NC quality, $p_j^{o\,\sublow}=b p_j^{o\,\subhigh}$. After substituting we discard the $\subhigh$ superscripts by redefining $p_j^\subhigh \equiv p_j$ and obtain the expression:

\begin{eqnarray}
s(t)& \propto & (k_2^{rad}-k_1^{rad})p_{j>1}^o\left(1+\frac{k_2}{k_3-k_2}\frac{p_{j>2}^o}{p_{j>1}^o}+\ldots\right) e^{-k_2 t} \nonumber \\
& & +\quad k_1^{rad}p_{j>0}^o\left(e^{-k_1 t}+b \cdot h(t)\right) \quad + \quad \ldots  \label{EQN_softmultiX}
\end{eqnarray}
where all $p_j$ and $k_j$ refer to the high quality $\subhigh$-type NCs. Eqn. \ref{EQN_softmultiX} can be used to interpret the results of our experimental data analysis. When studying the tPL decays, we first summarize the observed X dynamics by fitting to a biexponential $f(t)=e^{-k_{slow} t}+c_{fast} e^{-k_{fast} t}$, and then proceed to fit decays at higher excitation fluence to the form $s(t)=a_{BX}e^{-k_2 t}+a_{X}f(t)+(\ldots)$. Comparing to Eqn. \ref{EQN_softmultiX} we identify
\begin{equation}
a_{BX}=(k_2^{rad}-k_1^{rad})p_{j>1}^o\left(1+\frac{k_2}{k_3-k_2}\frac{p_{j>2}^o}{p_{j>1}^o}+\ldots\right)
\end{equation}
and
\begin{equation}
a_X=\left(\frac{1+b}{1+c_{fast}}\right)k_1^{rad} p_{j>0}^o 
\end{equation}
The first factor in this last expression is sensitive to details of sample subpopulations. One expects \emph{a priori} that $\frac{1+b}{1+c_{fast}}\sim 1$, which is the value it would take if X dynamics were fundamentally biexponential and were captured accurately in the fits (i.e. $k_1=k_{slow}$ and $h(t)=e^{-k_{fast} t}$), but there could be sample to sample variation to values smaller or larger than $1$. 


\subsection{Obtaining initial MX populations from tPL fits}

In studying CM in CdSe NCs, we are particularly interested in determining the fraction of photo-excited NCs that started in a BX state. This can be obtained from the ratio of fit components $a_{BX}/a_X$,
\begin{widetext}
\begin{equation}
\frac{a_{BX}}{a_X}=\underbrace{\frac{k_2^{rad}-k_1^{rad}}{k_1^{rad}}\left(\frac{1+b}{1+c_{fast}}\right)}_{\left(a_{BX}/a_X\right)_{sat}}
\frac{p_{j>1}^o}{p_{j>0}^o} \underbrace{\left(1+\frac{k_2}{k_3-k_2}\frac{p_{j>2}^o}{p_{j>1}^o}+\ldots\right)}_{1+\Delta}
\end{equation}
\end{widetext}

where we identify $(a_{BX}/a_X)_{sat}$ as the $a_{BX}/a_X$ ratio one would obtain if only the BX state were initially populated ($p_2^o=p_{j>0}^o$). $(a_{BX}/a_X)_{sat}$ can be estimated by fitting our expressions for $a_{BX}$ and $a_{X}$ using values extracted from an excitation power series. In an approximation, we carried out these fits neglecting the effect of the delay term in the $a_{BX}$ expression, effectively setting $ \Delta \approx 0$, which should lead to at most a $\approx 25 \%$ overestimate of $(a_{BX}/a_X)_{sat}$  considering the excitation fluences used.

	To determine CM yields with tPL, samples are excited with weak UV pulses and one looks for a BX signature in the resulting decay. Due to energy conservation, $p_{j>2}^o=0$ for our samples and excitation wavelengths. A simple expression is thus obtained for CM yield, $y_{cm}$, in terms of the fit parameters $a_{BX}$, $a_X$ and the $(a_{BX}/a_X)_{sat}$ value estimated from a previous excitation fluence power series:
	\begin{displaymath}
	y_{cm} \equiv \frac{p_2^o}{p_1^o+p_2^o}=\frac{a_{BX}}{a_X}\Bigg/\left(\frac{a_{BX}}{a_X}\right)_{sat}
	\end{displaymath}
	The method described for obtaining $y_{cm}$ is insensitive to the precise role of NC subpopulations in X and BX decays, and should be valid as long as $a_X\propto p_{j>0}^o$ and $a_{BX}\propto p_{j>1}^o$ under fairly general conditions. 
	
	On a final note, we see that one can estimate the ratio of BX to X radiative rates by:
	\begin{equation}
\frac{k_{BX}^{rad}}{k_X^{rad}}=1+\left(\frac{1+c_{fast}}{1+b}\right)
\left(\frac{a_{BX}}{a_X}\right)_{sat}\!\!\sim \, 1+\left(\frac{a_{BX}}{a_X}\right)_{sat} \nonumber
	\end{equation}
Since $b$ is not known independently, the $k_{BX}^{rad}/k_X^{rad}$ ratio thus obtained by assuming $b \approx c_{fast}$ has some uncertainty.

\subsection{Exact result for arbitrary $p_j^o$}
\label{SECTIONlaplace}
The system of kinetic equations (Eqn.\ref{EQNfirstorder}) can be solved for the components with dynamics proportional to $e^{-k_1 t}$ and $e^{-k_2 t}$ by Laplace transform. Define $P_j(s)=\mathcal{L}\{p_j(t)\}$.
Then
\begin{equation}
(s+k_j) P_j(s)=p_j^o+k_{j+1}P_{j+1}(s) \label{EQN_laplacediffeq}
\end{equation}
By iterating, one obtains the transformed solution.
\begin{eqnarray}
P_n(s)&=&\frac{1}{s+k_n}\left[p_n^o+\frac{k_{n+1}}{s+k_{n+1}}p_{n+1}^o \right. \nonumber \\
& &\quad+\left. \frac{k_{n+1}}{s+k_{n+1}}\frac{k_{n+2}}{s+k_{n+2}}p_{n+2}^o+\ldots \right] \label{EQN_laplacesolution}
\end{eqnarray}
which has a partial fraction decomposition of the form:
\begin{equation}
P_n(s)=\frac{A_n^{(n)}}{s+k_n}+\frac{A_{n+1}^{(n)}}{s+k_{n+1}}+\ldots=
\sum_{j\geq n} \frac{A_j^{(n)}}{s+k_j}
\nonumber
\end{equation}
Where the $A_j^{(n)}$ are constants independent of $s$. The solution $p_n(t)$ are thus given by
\begin{equation}
p_n(t)=A_n^{(n)}e^{-k_n t}+A_{n+1}^{(n)}e^{-k_{n+1} t}+\ldots=\sum_{j\geq n} A_j^{(n)}e^{-k_j t} \nonumber
\end{equation}
and the tPL signal is
\begin{equation}
s(t)=\sum_{n}k_n^{rad}p_n(t)=\sum_j e^{-k_j t}\underbrace{\left( \sum_{n\leq j}  k_n^{rad} A_j^{(n)} \right)}_{a_j}
\nonumber
\end{equation}
from which we are only interested in the $j=1$ (X) and $j=2$ (BX) components:
\begin{equation}
s(t)=\underbrace{k_1^{rad}A_1^{(1)}}_{a_X} e^{-k_1 t}+\underbrace{\left(k_2^{rad}A_2^{(2)}+k_1^{rad}A_2^{(1)}\right)}_{a_{BX}}
e^{-k_2 t} +\ldots
\nonumber
\end{equation}
Expressions for the $A_j^{(n)}$ coefficients involved are found by judicious application of the identity $\frac{1}{(s+a)(s+b)}=\frac{1}{b-a}\left(\frac{1}{s+a}-\frac{1}{s+b}\right)$ in Eqn. \ref{EQN_laplacesolution}. In particular,
\begin{widetext}
\begin{equation}
\label{EQN_Annsolution}
A_n^{(n)}=p_n^o+\left(1+\frac{k_{n}}{k_{n+1}-k_n}\right)p_{n+1}^o+
\left(1+\frac{k_{n}}{k_{n+1}-k_n}\right)\left(1+\frac{k_{n}}{k_{n+2}-k_n}\right)p_{n+2}^o+\ldots
\end{equation}
\end{widetext}
and it can be shown using Eqn. \ref{EQN_laplacediffeq} that $A_{n+1}^{(n)}=-\left(1+\frac{k_n}{k_{n+1}-k_n}\right)A_{n+1}^{(n+1)}$
This allows us to reduce the signal expression to
\begin{equation}
s(t)\approx\underbrace{k_1^{rad}A_1^{(1)}}_{a_X} e^{-k_1 t}+\underbrace{\left(k_2^{rad}-k_1^{rad}\right)A_2^{(2)}}_{a_{BX}}e^{-k_2 t}+\ldots \nonumber
\end{equation}
Where, using Eqn. \ref{EQN_Annsolution},
\begin{eqnarray}
A_1^{(1)}&=&p_1^o+\left(1+\frac{k_1}{k_2-k_1}\right)p_2^o+\ldots \nonumber \\
&\approx&p_1^o+p_2^o+p_3^o+\ldots\quad=\quad p_{j>0}^o \nonumber
\end{eqnarray}
In simplifying the above two equations we have again used the fact that the multiexciton decay rates are much faster than the X decay rate. Finally,
\begin{eqnarray}
A_2^{(2)}&=&p_2^o+\left(1+\frac{k_2}{k_3-k_2}\right)p_3^o\nonumber\\
& &\quad+\left(1+\frac{k_2}{k_3-k_2}\right)\left(1+\frac{k_2}{k_4-k_2}\right)p_4^o+\ldots \nonumber \\
&=&\left(p_2^o+p_3^o+\ldots\right)+\frac{k_2}{k_3-k_2}\left(p_3^o+p_4^o+\ldots\right)+\ldots \nonumber \\
&= & p_{j>1}^o\left(1+\frac{k_2}{k_3-k_2}\frac{p_{j>2}^o}{p_{j>1}^o}+\ldots \right)
\nonumber
\end{eqnarray}

which proves the general expressions for $a_X$ and $a_{BX}$ (Eqn. \ref{EQNax} \& \ref{EQNabx}).

\end{appendix}

\begin{acknowledgements}
	This work was funded in part by the NSF MRSEC program (DMR 0213282) at MIT and the authors made use its Shared Experimental Facilities.  It was also funded by the NSEC Program of the National Science Foundation (DMR-0117795), the David and Lucile Packard Foundation, the Department of Energy (DE-FG02-02ER45974), the Harrison Spectroscopy Laboratory (NSF-CHE-011370) and an NSF-NIRT (CHE-0507147). The authors would also like to thank Venda Porter, Scott Geyer, Numpon Insin, and Yinthai Chan for help in sample preparations.
\end{acknowledgements}


\begin{thebibliography}{29}
\bibitem{FOOTNOTE_NOFASTBX_FITDETAILS} 
At very high excitation fluence we do observe an additional very fast component but its dynamics appear consistent with $1S$-$1S$ emission from higher multiexcitonic states. The impact of this feature on further analysis was minimized by omitting early time ($t \alt \frac{1}{2}\tau_{BX}$) data from fitting.
\end{thebibliography}

\end{document}